\documentclass[%
floatfix,
 reprint,
 amsmath,amssymb,
 aps, pre
]{revtex4-2}
\usepackage[english]{babel}

\usepackage{amsfonts,array}
\usepackage{bm}
\usepackage{dcolumn}
\usepackage[output-decimal-marker={.}]{siunitx} 

\usepackage[final]{graphicx} 
\usepackage{listings} 
\usepackage[usenames]{color} 
\usepackage{color}
\usepackage{float}

\usepackage{hyperref}
\hypersetup{
    colorlinks=true,
    linkcolor={blue}, 
    filecolor=magenta,   
    citecolor={blue}, 
    urlcolor=blue, 
}

 \date{\today}

\newcommand{\br}{{\bf r}}
\newcommand{\bx}{{\bf x}} 
\newcommand{\bk}{{\bf k}} 
\newcommand{\by}{{\bf y}}

\begin{document}


\title{Impact of fluctuations on particle systems described by Dean-Kawasaki-type  
equations}

\author{Nathan O. Silvano}
\email{nathan@ifisc.uib-csic.es}
\author{Emilio Hern\'andez-Garc\'\i a}
\email{emilio@ifisc.uib-csic.es}
\author{Crist\'obal L\'opez}
\email{clopez@ifisc.uib-csic.es}
\affiliation{IFISC, Instituto de F\'isica Interdisciplinar
y Sistemas Complejos (CSIC-UIB), Campus Universitat de les
Illes Balears, E-07122 Palma de Mallorca, Spain}

\begin{abstract}
We study the role of fluctuations in particle systems modeled by 
Dean-Kawasaki-type equations, which describe the evolution of particle densities 
in systems with Brownian motion. By comparing microscopic simulations, stochastic 
partial differential equations, and their deterministic counterparts, we analyze 
four models of increasing complexity. Our results identify macroscopic quantities that can 
be altered by the conserved multiplicative noise that typically appears in the 
Dean-Kawasaki-type description. We find that 
this noise enhances front propagation speed in systems with density-dependent diffusivity, 
accelerates the onset of pattern formation in particle systems with nonlocal interactions, and reduces  
hysteresis in systems interacting via repulsive forces. In some cases, it 
accelerates transitions or induces structures absent in deterministic models. These 
findings illustrate that (conservative) fluctuations can have constructive and 
nontrivial effects, emphasizing the importance of stochastic modeling in understanding 
collective particle dynamics.    
\end{abstract}

\maketitle

\section{Introduction}
\label{Sec:intro}

Interacting Brownian particles are commonly used for modelling systems across a wide range of fields, from population ecology to condensed matter. The stochastic evolution of the particle density field in these systems, in the overdamped limit, is given by the so-called Dean-Kawasaki (DK) equation \cite{Dean1996,Kawasaki1994}, which is becoming an important tool in fields related to dynamic density functional theory and fluctuating hydrodynamics \cite{MarconiTarazona1999,Archer2004,Vrugt2020,Illien2025}.  The DK equation is a stochastic partial differential equation which models the dynamics of the density fluctuations, and which presents great challenges both analytically and numerically.  In recent years numerous studies have attempted to establish rigorous mathematical results and reliable numerical algorithms for it \cite{Cornalba2026,Djurdjevac2024,Konarovskyi2024}, or rather to extend the equation  to account for several species of different types  of particle mobilities, interactions or physical processes \cite{Nakamura2009,Poncet2017,Bressloff2024}. 


One of the characteristic features of the DK equation, and key source of its many difficulties, is the way it incorporates fluctuations. As it is a continuity equation expressing the conservation in the total number of particles, the fluctuation term appears in the form of conservative multiplicative noise. This involves the gradient of the square root of the density multiplied by Gaussian noise. In this paper we refer to {\sl DK-type equations} (DKTE) to
the ones that include this type of noise term. Unlike in non-conservative systems, where the noise term does not involve a gradient operator and for which there are now well-established numerical methods \cite{Dornic2005}, the conservative fluctuating term in the DK presents many difficulties, and the simplest approach is to neglect it (e.g. \cite{Delfau2016}), so that the role of fluctuations is  unknown. However, recent  studies \cite{Cornalba2026,Cornalba2023b} have proposed appropriate algorithms for the DK equation, showing \cite{Wehlitz2025} that they successfully reproduce the particle dynamics and analysing its range of validity. 
 
In this paper, we aim to study  the role of fluctuations in a series of particle models that can be described by DKTE. The models are presented in order of increasing  complexity, and we  analyse  relevant properties  like front propagation velocity, or the onset of pattern formation. Our analysis is performed from three perspectives: the particle model, its corresponding DKTE description (these two  are in principle equivalent), and the deterministic density equation, \textit{i.e}, the DKTE where the noise term has been neglected, hereafter referred to (somehow improperly) as deterministic DKTE. In this way we aim at understanding the influence of the conserved fluctuations. We adopt for the DKTE the numerical method from \cite{Cornalba2026,Cornalba2023b,Wehlitz2025}, which, as mentioned, provide rigorous statements of its validity. We also aim to test these types of algorithms as a function of the number of particles in the system, which is why we compare with the particle-based model. 
 
A crucial aspect of our study lies in the choice of models, which we consider in increasing order of complexity. This complexity  arises from the manner in which particles interact. With larger model complexity, in general we observe that the role of fluctuations becomes more significant. We begin by analyzing a system of non-interacting Brownian particles with a spatially dependent diffusivity. The steady solution is studied in the presence of a diffusivity gradient and the influence of fluctuations analysed.  Then we study again a system of Brownian particles but now with a density-dependent diffusion coefficient. This can be interpreted as some kind of local interaction among the individuals. For certain initial conditions, the system develops sharp fronts,
 whose velocity is examined. In contrast to other standard models where noise tends to slow down (pulled) front propagation \cite{Brunet1997}, we observe that in our case fluctuations enhance the front velocity. The third model considers a spatially non-local, density-dependent diffusion coefficient, so that the non-local interaction between the particles is reflected in the diffusivity. This system is known to exhibit spatial pattern formation \cite{Lopez2006}, with particles organizing, in two dimensions,
  into hexagonally ordered clusters. Again, we analyze the model across the three frameworks (particle, DKTE, and deterministic DKTE) and investigate the role of fluctuations in the onset and stability of these patterns. The last model we examine consists of soft-core repulsively interacting Brownian particles, a system that is also known to exhibit spatial patterns \cite{Delfau2016}. We study how conservative fluctuations affect the emergence and stability of these patterns. 

The outline of the paper is the following. In Section \ref{Sec:model1} we study the system of Brownian particles with a spatial gradient in the diffusivity. In Section \ref{Sec:model2} we study the density-dependent diffusivity system. Then in Section \ref{Sec:model3} we study the non-local version of this, and in Section \ref{Sec:model4} the model of repulsively interacting particles. Section \ref{Sec:summary} presents a summary and conclusions of our study.

\section{Brownian particles with spatially dependent diffusivity: Model I}
\label{Sec:model1}

Let us consider a onedimensional system, with size $L$, of $N$ non-interacting particles with positions $\{x_i\}_{i=1,...,N}$ which evolve as
\begin{equation} \frac{dx_i(t)}{dt}=\sqrt{2D (x_i) }\eta_i (t), \ \ i=1,..,N \,. \label{eq:model1_bugs} \end{equation} 
These equations are to be interpreted in the It\^o sense, with $\eta_i$ Gaussian 
white noises with zero mean, and correlations 
$\left< \eta_i (t) \eta_j (t') \right \rangle = \delta (t-t') \delta_{ij}$. 
Eq. (\ref{eq:model1_bugs}) 
can be derived \cite{Lopez2007,Celani2012} for the positions of a set of noninteracting Brownian 
particles in the overdamped 
limit with a constant friction coefficient $\gamma$ but in an inhomogeneous  
temperature field $T(x)$ which reflects in the position-dependent diffusion coefficient 
$D(x)=k_B T(x)/\gamma$. Alternatively, here we consider Eq. (\ref{eq:model1_bugs}) as 
the simplest example within a hierarchy of stochastic models describing the 
motion of physical or biological point entities that will be used in this paper 
to illustrate the impact of fluctuations on particle densities and its proper modelling. 

 
We first derive the DKTE following closely the approach in \cite{Dean1996}. As usual we define the single particle density $\rho_i (x,t)= \delta (x_i(t)-x)$, and consider an arbitrary function $f$ so that
\begin{align}
f(x_i)=\int dx\, \delta (x_i-x) f(x) =\int dx \rho_i (x,t) f(x)\,.
\end{align}
Following the calculation by Dean using It\^o lemma:    
\begin{eqnarray}
\frac{df (x_i)}{dt} &= &\frac{\partial f (x_i)}{\partial x_i} \frac{dx_i}{dt} 
+ \frac{1}{2} 2 D(x_i) \frac{\partial^2 f}{\partial x_i^2}  \nonumber \\
&=&\frac{\partial f (x_i)}{\partial x_i} \sqrt{2D (x_i(t)) }\eta_i (t) + 
D(x_i) \frac{\partial ^2 f}{\partial x_i^2} \nonumber \\
&=&\int dx \rho_i (x,t) \left[ \frac{\partial f (x)}{\partial x} 
\sqrt{2D(x) }\eta_i (t) +  D(x) \frac{\partial^2 f}{\partial x^2} \right].
\end{eqnarray}

On the other hand
$\frac{df (x_i)}{dt}
=\int dx \partial_t \rho_i (x,t) f(x)$.
Integrating by parts and equating integrands:
\begin{equation}
\partial_t \rho_i (x,t) = \frac{\partial^2}{\partial x^2} ( D (x) \rho_i (x,t)) 
- \frac{\partial}{\partial x}(\sqrt{2D(x) } \rho_i \eta_i).
\end{equation}

Summing up over all the particles  and defining the total density of particles 
$\rho (x,t)=\sum_{i=1}^N \rho_i (x,t)= 
\sum_{i=1}^N\delta (x_i(t)-x)$:
\begin{equation}
\partial_t \rho (x,t) = \frac{\partial^2}{\partial x^2} 
( D (x) \rho (x,t)) - \frac{\partial}{\partial x} (\sqrt{2D(x) } \sum_i \rho_i \eta_i) \ .
\end{equation}
It is standard to show, by analyzing the statistical properties of the last noise term
\cite{Dean1996}, that this is equivalent to the following DKTE for
the total density:
\begin{equation}
\partial_t \rho (x,t) = \frac{\partial^2}{\partial x^2} \left(D (x) \rho (x,t)\right) 
+ \frac{\partial}{\partial x} \left(\sqrt{2D(x) \rho (x)} \zeta (x,t)\right),
\label{DKmodel1}
\end{equation}
where $\left< \zeta(x,t) \zeta (x',t') \right> = \delta (x-x') \delta (t-t')$.
This expression is better analysed in its
non-dimensional form. For this we define the
non-dimensional variables $\tilde x = x/L$,
$\tilde t= t D_0/L^2$ and $\tilde \rho = \rho/\rho_0$,
where $\rho_0= N/L$ is the initial density (which
is conserved) and $D_0$ is a constant with the dimensions
of a diffusion coefficient. Omitting the tilde notation
in most of the variables we obtain:
\begin{equation}
\partial_t \rho (x,t) = \frac{\partial^2}{\partial x^2} 
\left(\tilde D (x) \rho (x,t)\right) 
+ \frac{1}{\sqrt{N}}\frac{\partial}{\partial x} 
\left(\sqrt{2 \tilde D(x) \rho (x)} \zeta (x,t) \right),
\label{DKmodel1adim}
\end{equation}
where $\tilde D (x)= D(x)/D_0$. Here it is explicitly seen that in the limit $N\to \infty$ we get a deterministic description. Note that the normalization of the scaled density appearing in (\ref{DKmodel1adim}) is such that $\int dx\rho(x,t)=1$. Note also that Eq.(\ref{DKmodel1}) (or Eq.(\ref{DKmodel1adim})) is completely equivalent to the particle dynamics described by Eq.(\ref{eq:model1_bugs}). However, our objective is to compare the particle dynamics with the DKTE as numerically approximated, and both to the deterministic version of the DKTE, in order to uncover the role of fluctuations on some of the properties of this model.

\subsection{Numerical results}
\label{Sub:model1}

The numerical simulation of the DKTE is based on the algorithm proposed in \cite{Cornalba2026,Cornalba2023b,Wehlitz2025}. These studies demonstrate that a standard discretization of the DK equation, using either finite difference or finite element schemes, provides a suitable approximation for capturing density fluctuations in a system of $N$ diffusing particles at sufficiently fine discretization, provided that at each time step and at every point the density appearing in the noise term is replaced by the maximum between zero and the density computed there from the previous step, thus preventing the accumulation of unphysical negative densities. This numerical approximation works better in a high density regime since one of it sources of error is proportional to $\exp(-N h^d)$, where $h$ is the numerical spatial grid size and $d$ the spatial dimension. Note that, since $h$ here is measured in units of system size, $N h^d$ is equal to $\rho_0$ times the unscaled grid volume. Thus, the small error condition means that  as long as there is substantially more than one particle per grid cell (\(Nh^d \gg 1\)) the numerically discretized density  accurately describes the particle system. The details of the numerical integration of this model and of the following ones can be found in the Appendix. 
 
\begin{figure}[hbt!]
	\begin{center}
		\includegraphics[width=\columnwidth]{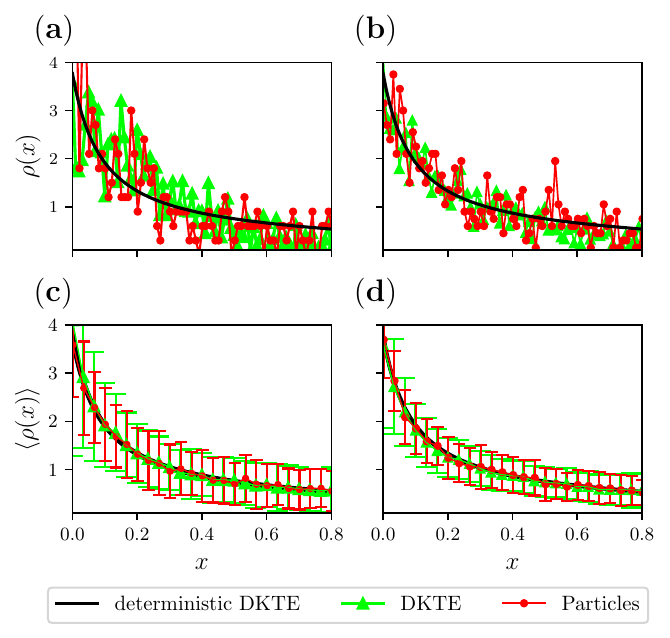}
\caption{Density distributions measured at long times, when statistically steady states are reached, for model I in the $[0,1]$ interval (only a slightly smaller region is shown). In all panels, the black line is the exact solution of the deterministic DKTE, corresponding to $N \to \infty$, green triangles are from the DKTE and red circles from the particle simulations. For the particles, densities are obtained by binning the particle positions in histograms of 300 bins. Instantaneous configurations are displayed in panels (\textbf{a}) and (\textbf{b}), whereas their averages over long times are in panels (\textbf{c}) and (\textbf{d}). There, the error bars indicate the standard deviation of the density fluctuations at each point.  (\textbf{a}) and (\textbf{c}) are for $N=1000$ and (\textbf{b}) and (\textbf{d}) for $N=2000$. We note the very good agreement between the variances of the particle and DKTE simulations and of the average densities between them and with the deterministic solution. Simulations of DKTE were performed with $dt = 10^{-6}$ and $dx =3.33\times 10^{-3}$, up to $5\times 10^{6} $ simulation steps, while for the particles $dt = 10^{-6}$ with $10^6$ steps, and we binned the particle positions into a density histogram with the same $dx$ as the DKTE.}
\label{Fig:model0}
\end{center}
\end{figure}

For the numerical example we consider a domain $[0,L]$. We
impose zero-flux conditions at $x=0$ and $x=L$.
For the spatially-dependent diffusivity we take:  
\begin{align}
\tilde D(x) = 0.1 +  x \ ,
\end{align}
and $L=1$.  The exact solution of the deterministic model I with zero-flux boundary conditions is $\rho(x)=C/\tilde D(x)$, with $C$ a normalization constant.  In Fig.\,\ref{Fig:model0} we present the steady density profiles for the particle dynamics, the DKTE for two values of $N$,  and the deterministic DK (\textit{i.e}. $N\to \infty$). The only effect of fluctuations is the emergence of a rough density profile (equivalent for DKTE and particles) relative to the smooth  deterministic solution.  This is further illustrated when we consider the average density over realizations $\langle \rho \rangle$ which evolves according to  $ \partial_t \langle \rho (x,t) \rangle = \frac{\partial^2}{\partial x^2} (\tilde D (x) \langle \rho (x,t)\rangle)$. Thus, this equation for the mean field is identical to the  deterministic DKTE given by $N\to \infty$. The lower panels of Fig.\,\ref{Fig:model0} display the long-time average of the density $\left< \rho \right>$ for the DKTE and particles, that compare perfectly with the deterministic DKTE description. The impact of fluctuations on the steady density is also independent of the function $\tilde D(x)$. 
 
\vspace{0.5cm}


The lower panels of Fig.\,\ref{Fig:model0} also display the standard deviations obtained from the particle simulations and the DKTE. The good agreement confirms that the numerical method used for the DKTE description accurately 
captures the particle dynamics.

We remark that there is no explicit coarse graining \cite{Archer2004,uneyama2007} in the DKTE simulations presented here and in the following sections, as they are intended to approach the microscopic particle density $\rho(x,t)=\sum_{i=1}^N \delta(x_i(t)-x)$. But, of course, such a singular distribution can not be exactly reproduced in a finite-resolution numerical simulation, so that an implicit regularization of it is performed by the numerical algorithm. The mathematical results in \cite{Cornalba2026,Cornalba2023b} show that as the resolution in our numerical algorithm improves, the simulated (regularized) density $\rho(x,t)$ approaches the one from the particle model in a well-defined weak-limit sense. For a finite resolution, and in order to compare the results from DKTE and particle simulations, here and in the following the microscopic densities from the particle simulations are converted into histograms by binning to the same spatial discretization step as in the DKTE numerics, unless otherwise stated. 

\section{Brownian particles with density-dependent diffusivity: Model II}
\label{Sec:model2}

In the next model, also  in one spatial dimension, we consider diffusivity $D=D(\rho)$ to be density-dependent, that is, it varies with the number of particles at each point in space
(we refer to this as model II). Density-dependent diffusion can be seen as an effective way of considering interparticle interactions (e.g. \cite{Tailleur2008}). A possible formal description at the particle level is given by:
\begin{equation}
\frac{dx_i(t)}{dt}=\sqrt{2D \left( \sum_{j=1}^N \delta(x_i (t) -x_j(t))\right) }\eta_i (t), \ \ i=1,...,N.
\label{model2}
\end{equation}

Since we have 
\begin{eqnarray}
\sum_{j=1}^N \delta(x_i (t) -x_j(t))& = &\sum_{j=1}^N \int 
dy \delta(x_i-y) \delta(y-x_j) \nonumber \nonumber \nonumber \\
& = &\int dy \delta(x_i-y) \sum_{j=1}^N \delta(y-x_j) \nonumber \\
 & = &\int dy \delta(x_i-y) \rho (y,t) \nonumber \\ &=& \rho(x_i,t),
\label{xmodel2}
\end{eqnarray}
we can rewrite Eq. (\ref{model2}) in a more illustrative form as
\begin{equation}
\frac{dx_i(t)}{dt}=\sqrt{2D (\rho (x_i,t)) }\eta_i (t), \ \ i=1,...,N.
\label{model2b}
\end{equation}
The DKTE for this system of particles is obtained 
following similar steps as before:
\begin{align}
\frac{df (x_i)}{dt} &=\frac{\partial f (x_i)}{\partial x_i} \frac{dx_i}{dt} +
 D(\rho(x_i)) \frac{\partial^2 f}{\partial x_i^2} \nonumber \\
&=\frac{\partial f (x_i)}{\partial x_i} \sqrt{2D (\rho (x_i)) }\eta_i (t) + 
 D(\rho(x_i)) \frac{\partial ^2 f}{\partial x_i^2} \nonumber \\
=&\int dy \rho_i (y,t) \left[\frac{\partial f (y)}{\partial y}
 \sqrt{2D(\rho(y)) }\eta_i (t)  
 +  D(\rho(y)) \frac{\partial^2 f}{\partial y^2} \right] \nonumber \\
&=\int dy \partial_t \rho_i (y,t) f(y) \ .
\end{align}
Integrating by parts, equating integrands and summing over all particles,
the final expression for the density is similar to the model before
but now with $D(\rho)$:
\begin{equation}
\partial_t \rho (x,t) = \frac{\partial^2}{\partial x^2} 
( D (\rho) \rho (y,t)) +\frac{\partial}{\partial x} (\sqrt{2D(\rho) \rho (x)} \zeta (x,t)) \ .
\label{DK2}
\end{equation}
The deterministic versions of this type of equations are generically known as nonlinear diffusion equations \cite{Logan}.

In the next subsection we  compare, numerically and with the same algorithm specified in the previous section, the different approximations for this model, \textit{i.e}., the particle dynamics Eq.(\ref{model2b}), 
the corresponding DKTE, Eq.(\ref{DK2}), including an analysis of the average (over realizations) density $\langle \rho \rangle$, and the deterministic DKTE.

\subsection{Numerical results.}
\label{Sub:model2}

To proceed we first assume a specific form for the diffusivity $D(\rho)= D_0 (\frac{\rho}{\rho_0})^n$, where $D_0$ and $n$ are positive constants. This type of density dependence appears naturally in collections of individuals (like insects) that enhance their motility due to population pressure \cite{Murray}. By introducing the nondimensional variables $\tilde x =x/L$, $\tilde t = t D_0/L^2$, $\tilde \rho = \rho/\rho_0$ and omitting  the tilde notation for simplicity, we obtain:
\begin{equation}
\partial_t \rho (x,t) = \frac{\partial^2}{\partial x^2} (\rho(x,t)^{n+1} ) +
\frac{1}{\sqrt{N}}\frac{\partial}{\partial x} (\sqrt{2 \rho^{n+1}} \zeta (x,t)),
\label{DK2adim}
\end{equation}
where now the dimensionless diffusivity is 
$D= \rho^n$ and the noise
correlations $\left<\zeta (x,t) \zeta (x',t')\right> = \delta (x - x') \delta (t - t')$.

\begin{figure}[hbt!]
\begin{center}
\includegraphics[width=0.9\columnwidth]{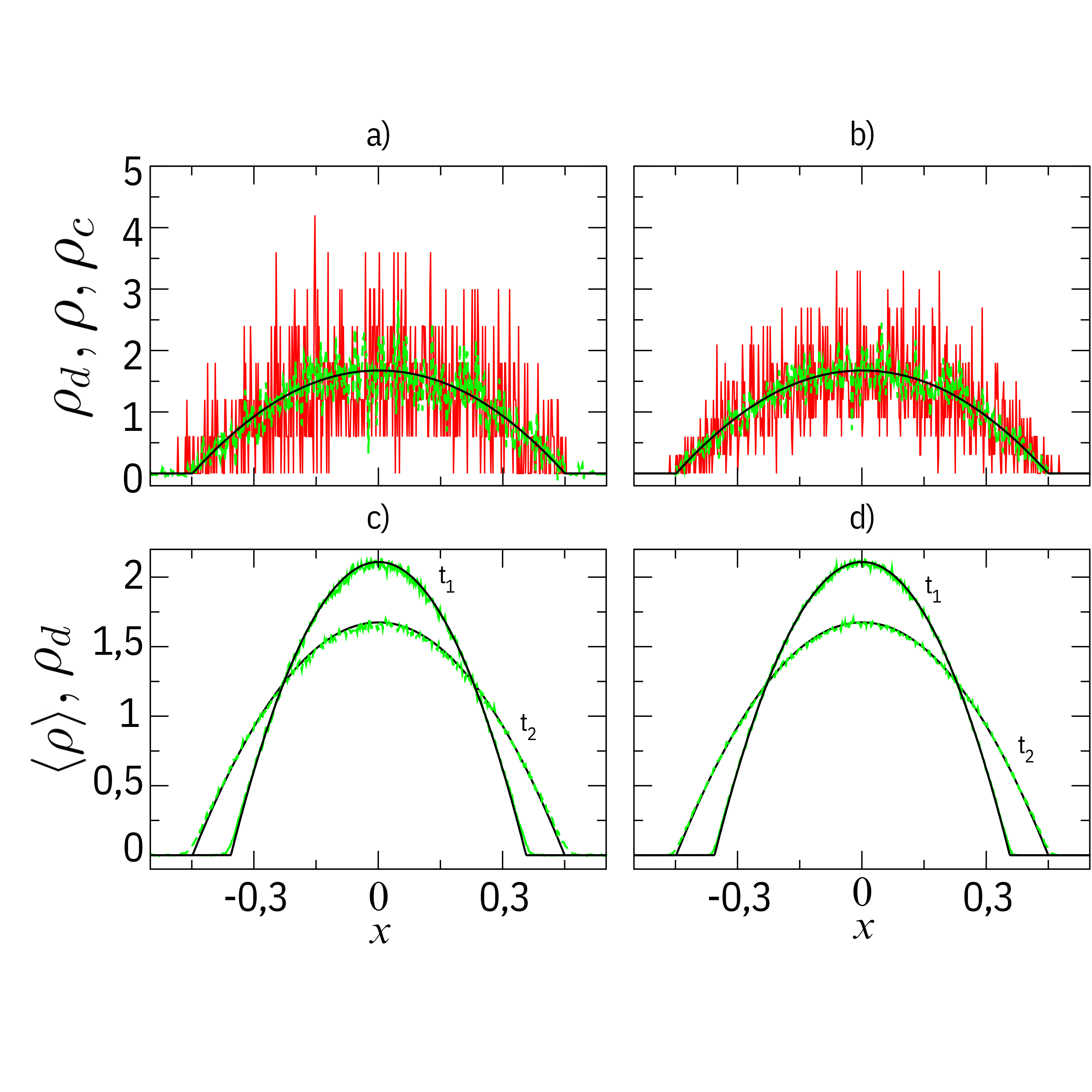}
\caption{Model II: \textbf{a}) $N=1000$. Black line: exact solution, $\rho_d$,  
of the deterministic DKTE at time $t=0.01$. Green line: numerical solution, $\rho$, of the DKTE, Eq. (\ref{DK2adim}), at the same time. Red line: 
density $\rho_c$ obtained from particle simulations of Eq.(\ref{model2b}). 
\textbf{b}) the same for $N=2000$ particles. \textbf{c}) $N=1000$. 
 Black line is again the 
analytic solution of the deterministic DKTE at two  different times $t_1=0.005, t_2=0.01$. Green line is the average over $300$ realizations, all
for the same initial condition, of the solution of the DKTE 
at the same times. \textbf{d}) The same as \textbf{c}) for $N=2000$.
The numerical simulations for DKTE (see Appendix) are performed with
$dt=10^{-8}$ and $dx=0.00167$, particle simulations with 
$dt=10^{-4}$, and the resulting particle distribution binned to the same $dx$ as the DKTE. 
}
\label{Fig:DK2Sol}
\end{center}
\end{figure}

The nonlinear diffusion equation, represented by 
the deterministic version of Eq.(\ref{DK2adim}), $\partial_t \rho_d (x,t) = \frac{\partial^2}{\partial x^2} (\rho_d (x,t)^{n+1})$, has been extensively studied \cite{Logan}.
In particular for $n=1$ and any compact-support initial condition, the solution approaches the parabolic form $\rho_d (x,t) = (A^2 t^{2/3} - x^2)/(12t)$ for $|x| < A t^{1/3}$, and $\rho_d (x,t) =0$ elsewhere. This represents a hump of particles limited by two sharp fronts moving in opposite directions at positions $x_f(t)=\pm  A t^{1/3}$ ($A$ is a positive constant determined by normalization). We will focus our analysis of this model in this type of sharp front solutions. We perform simulations starting from an initial condition that is already of the above analytical form
(i.e. we take as initial condition $\rho_d (x,t=10^{-5})$).
  See the Appendix for the details of the numerical method. 
  The upper panels of Fig.\,\ref{Fig:DK2Sol} display the 
  solutions $\rho(x,t)$ of the DKTE, Eq. (\ref{DK2adim}), 
  at a later time ($t=0.01$). The analytic solution $\rho_d(x,t)$ from 
  the deterministic DKTE, and particle densities, $\rho_c$,
   constructed from histograms of their positions are also shown. 
A slight difference between the stochastic 
and deterministic solutions is observed in 
their temporal evolution since the front of 
the stochastic solution always seems to advance the 
deterministic one. This is more clearly seen in the
bottom plots of Fig.\,\ref{Fig:DK2Sol} where we plot
the same deterministic DKTE solution and the ensemble average $\left< \rho \right>$ 
over $300$ realizations of the DKTE, all from the 
same initial condition, at two different times ($t=0.005$ and $t=0.01$) 
to see the temporal evolution. 

\vspace{0.5cm}

\begin{figure}[hbt!]
\begin{center}
\includegraphics[width=0.9\columnwidth]{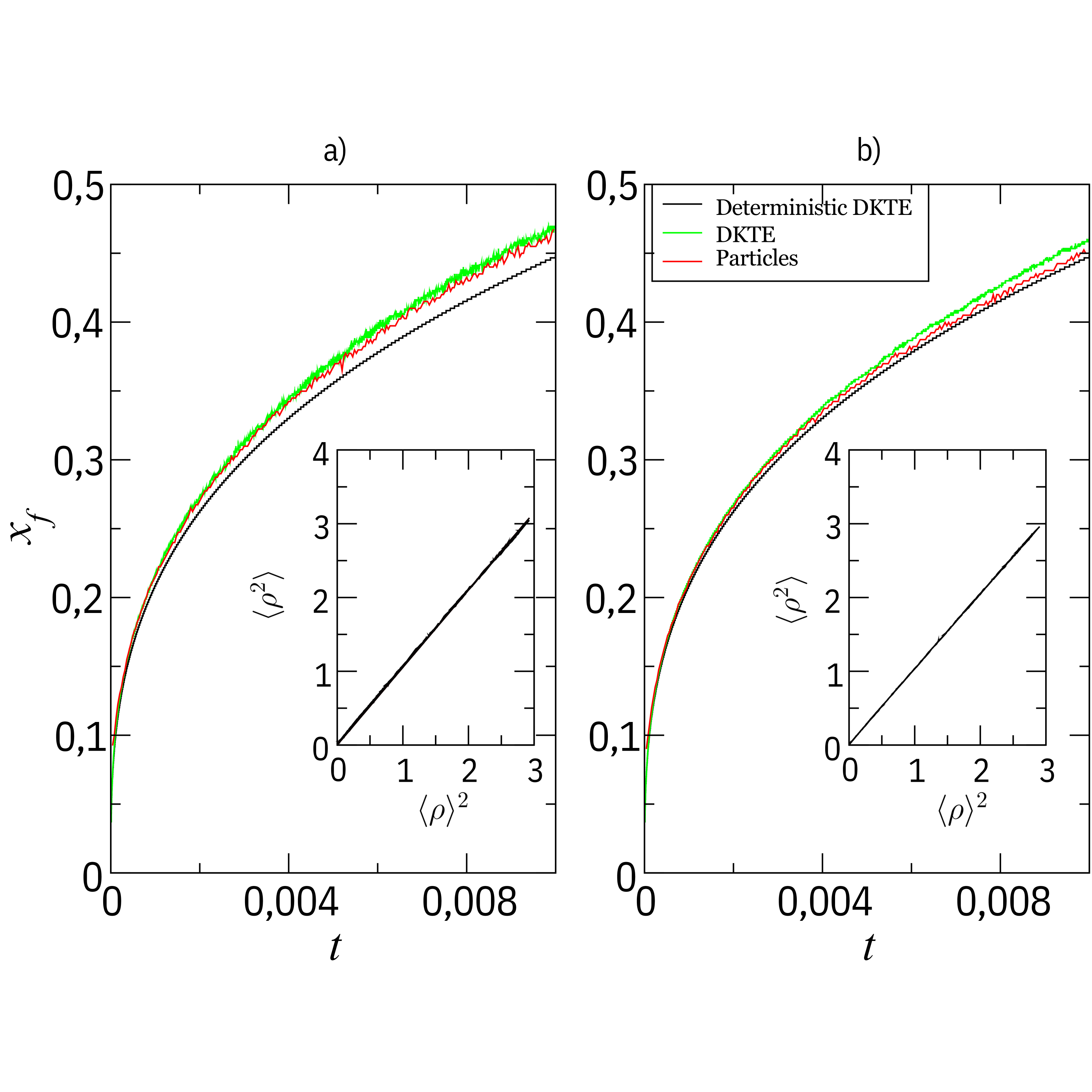}
\caption{Model II: 
Front positions vs time for the deterministic equation (black line), 
the average over $300$ realizations of the DKTE (green line), and the average over $300$ realizations of the density obtained from position histograms of particle simulations (red). The front is located as the rightmost spatial point for which the density is equal or larger than $0.01$. \textbf{a}) is for $N=1000$, \textbf{b}) is for $N=2000$. Other parameters as in Fig. (\ref{Fig:DK2Sol}). The insets show the relation between $\left< \rho^2 \right>$ and $\left< \rho \right>^2$ at time $t=0.01$ at the different spatial points.  
}
\label{Fig:fronts}
\end{center}
\end{figure}

To better quantify these results, we compute
the position of one of the sharp fronts
(the right one), $x_f(t)$, as a function of time, 
for all the frameworks previously discussed: 
the particle simulations (computing a density 
field via an histogram, see caption of Fig.\,\ref{Fig:DK2Sol}), 
the numerical DKTE solution $\rho (x,t)$, and the analytic solution $\rho_d (x,t)$ of the deterministic DKTE. 
We compute the front  for the ensemble 
average of the density, $\langle\rho\rangle$, 
over $300$ realizations for DKTE 
and particles (the average is not needed 
for the deterministic DKTE). 
The front position is located as the rightmost point
 for which density exceeds a threshold equal to $0.01$
 The results are shown in Fig.\,\ref{Fig:fronts}. The behavior of the front positions as a function of time are all well fitted with the same power law $x_f(t) \propto t^{1/3}$, implying that the fronts in the average density of the DKTE or the particles effectively behaves as in the deterministic case, but with a larger diffusivity   
 (if a diffusivity different from unity is used in Eq.(\ref{DK2adim}) the prefactor $A$ in the front position becomes an increasing function of this diffusivity).

\vspace{0.5cm}

An heuristic way to understand this result can be obtained by taking
the ensemble average of Eq.(\ref{DK2adim}) (with $n=1$), which leads to  
(the noise disappears under averaging since
we use It\^o Calculus):
\begin{equation}
\partial_t \left< \rho (x,t) \right> = \frac{\partial^2}{\partial x^2} (\left< \rho(x,t)^2 \right> ).
\label{DK2calc1}
\end{equation}

Numerically, we observe (insets of Fig. (\ref{Fig:fronts})) a rather accurate, which seems
to not vary with time, linear relationship between $\left< \rho^2 \right>$ and $\left< \rho \right>$: $\left< \rho^2 \right> \approx \hat D \left< \rho \right>^2$, with $\hat D= 1.037$ (for $N=1000$) and $\hat D=1.018$ (for $N=2000$). Substituting into Eq. (\ref{DK2calc1}) we see that the dynamics of the average density is ruled by an equation of the same form as the deterministic DKTE but with a larger effective diffusion coefficient $\hat D > 1$: 
\begin{equation}
\partial_t \left< \rho (x,t) \right> \approx \hat D \frac{\partial^2}{\partial x^2} (\left< \rho(x,t)\right>^2 ),
\label{DK2calc3}
\end{equation}
which implies that the fronts in the average density of the DKTE move with the same exponent $t^{1/3}$ as the deterministic ones, but with a larger prefactor arising from the larger diffusion coefficient $\hat D$, which is a consequence of the fluctuations.

It is worth emphasizing that, in this model described by DKTE, the velocity of the sharp front is actually enhanced. This contrasts with standard pulled fronts in noisy (non-conserved) FKPP equations, where fluctuations typically act to slow down the front propagation \cite{Brunet1997}. Those studies have also analysed the dependence of the front velocity with $N$, showing that corrections to the deterministic front velocity scale as $1/log^2 N$. The dependence on $N$ of the sharp front dynamics for our model is left to future work.

\section{Brownian particles with nonlocal density-dependent diffusivity: Model III}
\label{Sec:model3}

Continuing with our hierarchy of models, we now consider an extension of the previous one in which the diffusion coefficient depends non-locally on the density, \textit{i.e}., $D=D(\rho_G ({\bf x},t))$, where $\rho_G ({\bf x},t) = \int d{\bf y} G({\bf x}-{\bf y}) \rho ({\bf y},t)$ and the kernel $G({\bf x}-{\bf y})$ quantifies the influence of density at ${\bf y}$ on location ${\bf x}$. Since we  focus this study on the spatial structures that form, the system will be considered in two dimensions. The simplest case corresponds to a top-hat kernel: $G({\bf x})=1$ if $|\bx|\leq R$ and zero otherwise. Under this choice $\rho_G  ({\bf x},t) = N_R ({\bf x},t)$ becomes the number of particles in a radius $R$ around ${\bf x}$ at time $t$. 

At a particle level, this non-local dependence of the diffusion coefficient on the density is described by the following equations for
the $i=1,...,N$ particles:
\begin{equation}
\frac{d\bx_i(t)}{dt}=\sqrt{2D\left(\sum_{k : |\bx_i-\bx_k|\leq R} \ \sum_{j=1}^N \delta(\bx_k (t) -\bx_j (t))\right)}\boldsymbol\eta_i (t).
\label{model3}
\end{equation}
$\boldsymbol\eta_i (t)$ is a Gaussian white-noise vector acting on particle $i$, with zero mean and correlations 
$\left<\boldsymbol\eta_i (t)\boldsymbol\eta_j (t')\right>=\mathbb{I} \delta_{ij}\delta(t-t') $. $\mathbb{I}$ is the $2\times 2$ identity matrix. Again we have
\begin{eqnarray}
\sum_{k : |\bx_i-\bx_k|\leq R} \ \sum_{j=1}^N \delta(\bx_k (t) -\bx_j(t)) &=& \sum_{k: |\bx_i-\bx_k|\leq R} \ \rho(\bx_k,t)
\nonumber \\ &=&\rho_G(\bx_i,t)
\end{eqnarray}
so that in the continuum notation, $\rho_G(\bx_i,t) =\int_{|\bx_i-\br|\leq R} d\br \rho(\br,t)$,
and we have $D=D(\rho_G(\bx_i,t))$. Thus, we can write the particle equations as 
\begin{equation}
	\frac{d\bx_i(t)}{dt}=\sqrt{2D(\rho_G (\bx_i,t))} ~\boldsymbol\eta_i (t).
\label{model3ill}
\end{equation}

Proceeding similarly to the previous section we obtain the corresponding DKTE:
\begin{equation}
\partial_t \rho (\bx,t) = \boldsymbol\nabla^2 ( D (\rho_G) \rho (\bx,t)) 
+\boldsymbol\nabla \cdot (\sqrt{2D(\rho_G) \rho (\bx)} \boldsymbol\zeta (\bx,t)), 
\end{equation}
where now $\boldsymbol\zeta (\bx,t)$ is a spatiotemporal Gaussian white-noise vector with zero mean and correlations 
$\left<\boldsymbol\zeta (\bx,t)\boldsymbol\zeta (\bx',t')\right>=\mathbb{I} \delta(\bx-\bx')\delta(t-t') $. 

\subsection{Numerical results}
\label{Sub:model3}

As in previous cases, we analyze the influence of fluctuations in the model with spatially non-local, density-dependent diffusivity. To proceed, we assume a specific form for the diffusivity functional and, similarly to the previous model, we take $D(\rho_G)= D_0 ( \rho_G/(L^2\rho_0))^p$, where $D_0$ is a positive constant with dimensions of diffusivity. The nondimensional version of the DKTE is now
\begin{equation}
\partial_t \rho (\br,t) = \boldsymbol\nabla^2 ((\rho_G)^p \rho) +\frac{1}{\sqrt{N}} \boldsymbol\nabla \cdot (\sqrt{2 ( \rho_G )^p \rho} ~\boldsymbol\zeta (\br,t)).
\label{eq:DKmodelIII}
\end{equation}

Ref. \cite{Lopez2006} studied this type of dynamics both neglecting fluctuations and in an equivalent particle system. In both cases, the most relevant feature was that, in two spatial dimensions, density shows spatially periodic cluster patterns of hexagonal symmetry 
if $p$ is large enough. Our focus here is to analyze, in two dimensions, this behavior in the framework of the DKTE, and compare it with the particle and the deterministic descriptions. To this end we perform numerical simulations of the three frameworks. Numerical details are in the Appendix. 

In Fig.\,\ref{fig:model3snap} we show long-time density distributions from this model III. For $p= 6.5$  we have statistically homogeneous distributions, while for $p =8.0$ spatially periodic cluster patterns develop in the DKTE, its deterministic version, and in the particle simulations.

The characterization of the emerging spatially periodic patterns is carried out using the structure function, whose peaks at particular wavenumbers indicate the presence of periodic structures with that periodicity.  The structure function for the continuum system is computed as the modulus of the Fourier transform of the steady-state normalized density and then averaged spherically and over time. For particles it is computed as $\left< |\sum_i |e^{\bk \cdot \bx_i (t)}|^2/N^2 \right>$, where $\bx_i (t)$ is the position of particle $i$, $\bk$ is a twodimensional wave vector, and the average indicates a spherical average over the wave vectors with modulus $k$, a temporal average in the steady state, and an average over many different realizations. Well-developed spatially periodic patterns of period $l$ are associated with pronounced peaks in the structure function at wavenumbers $|{\bf k}|=2\pi/l$. 

In Fig.\,\ref{fig:model3} we plot the height of the peak of the structure function in terms of the parameter $p$ for systems of $N=10000$ and $N=20000$ particles (remember this is a 2d system at difference of the previous sections). The transitions are observed at a critical value of $p_c$ which is $7.6$ for the deterministic case, $7.3$ for the particles and $6.8$ for the DKTE. The value of $p_c$ for the deterministic case coincides with the one calculated via linear stability analysis of the homogeneous solution \cite{Lopez2006}. 

A salient feature of this figure is the presence of a small region of noise-induced patterns \cite{OjalvoSanchobook}, \textit{i.e.}, a region of $p$ values smaller than the one needed to produce periodic patterns in the deterministic case, but where fluctuations give rise to noisy but well-developed spatial periodicity. In both the particle simulations and the DKTE, the transition to pattern formation occurs for values of the control parameter $p$ smaller than in the deterministic case.
The small discrepancies between DKTE and particles simulations are
due to the numerical details in the first.
Additional features, such as the presence of a hysteresis loop (which is not shown here),
 will be discussed in detail in the next section in the context of a different model.

\begin{figure}[hbt!]
	\begin{center}
		\includegraphics[width=\columnwidth]{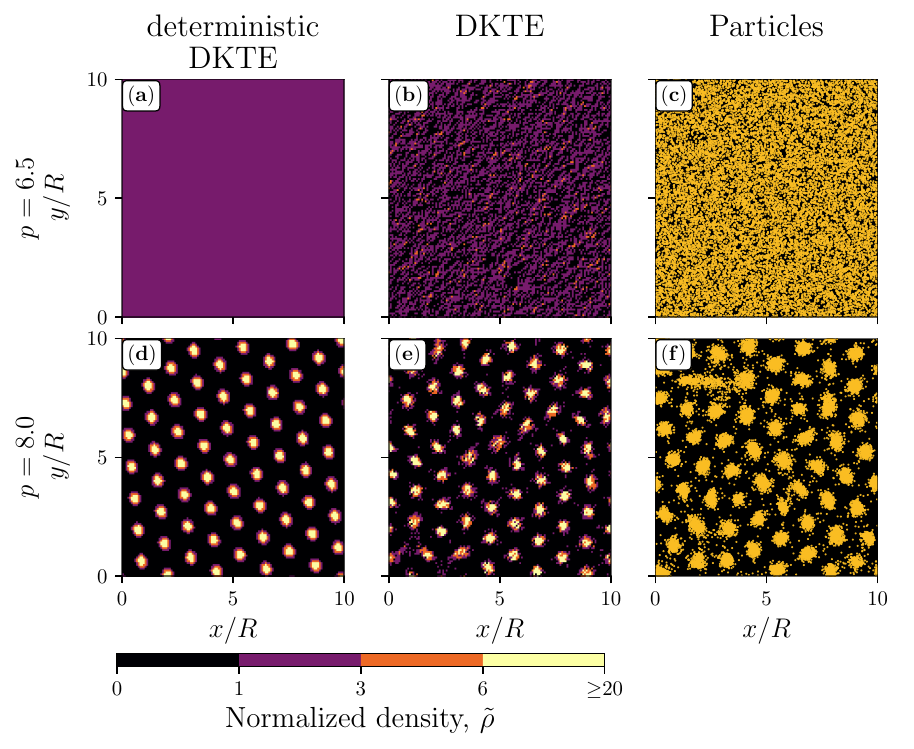}
		\caption{Normalized density distributions, $\tilde\rho=\rho/\rho_0$, from model III at long times and two different values of $p$. \textbf{(a-c)} are statistically homogeneous distributions with $p = 6.5$. \textbf{(d-f)} are statistically steady density patterns for $p= 8.0$. The deterministic DKTE (\textbf{a,d}) and the DKTE (\textbf{b,e}) profiles were obtained from the simulation of Eq.~\eqref{eq:DKmodelIII} for $L=1$, $R=0.1$, with $dt = 10^{-4}, dx =7.8\times 10^{-3}$, and $N= \infty$ and $N=20000$, respectively. Particle data (\textbf{c,f}) where obtained from simulations of Eq.~\eqref{model3ill} for $N = 20000$, $dt = 10^{-1}$ and $10^8$ steps. The Particle images are shown without any spatial binning, i.e. each yellow dot indicates the position of a particle.}
		\label{fig:model3snap}
	\end{center}
\end{figure}

\vspace{0.5cm}
\begin{figure}[hbt!]
\begin{center}
\includegraphics[width=\columnwidth]{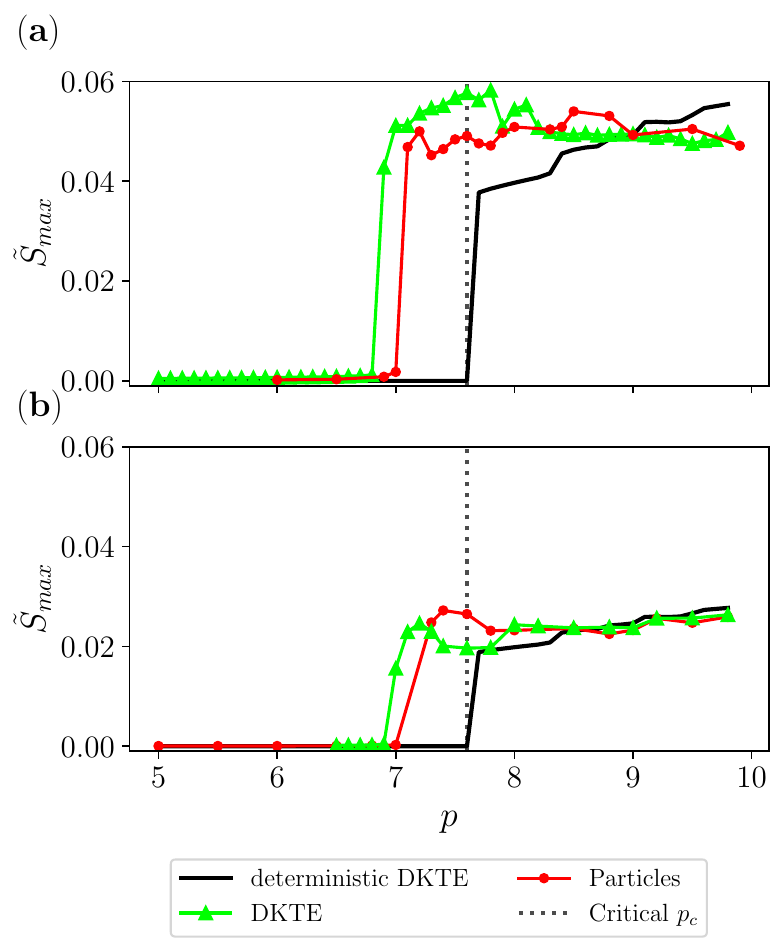}
\caption{Height of the peak of the structure function, normalized 
with $N$, as a function of the parameter
$p$ for the three frameworks: deterministic DKTE (black), DKTE (green), and particle (red). 
\textbf{a}) is for $N=10000$ and \textbf{b}) for $N=20000$. The deterministic DKTE is integrated up to $1.5\times 10^{6}$, and the DKTE up to  $2.5\times 10^{6}$ simulation steps. 
Other parameters as in Fig. \ref{fig:model3snap}. The vertical dotted line indicates the 
value of $p_c$ identified by
linear stability analysis of the
homogeneous solution
in the deterministic case.
}
\label{fig:model3}
\end{center}
\end{figure}

\section{Repulsively interacting Brownian particles: Model IV }
\label{Sec:model4}
 
The last system we consider is a collection of Brownian particles interacting through a pair potential in a two-dimensional domain. In contrast to the previous models, where particle interactions were mediated indirectly through the diffusivity, the current model considers a direct inter-particle force. Thus, in this case the DKTE is in fact the original DK equation \cite{Dean1996,Kawasaki1994}.  This system has been extensively studied in \cite{Delfau2016} in the twodimensional case with repulsive soft-core potentials, with the somehow surprising result that, despite the repulsive nature of the interactions, particles can spontaneously aggregate forming a periodic crystal of particle clusters. The resulting spatial pattern is very similar to the one obtained with the model of the previous section (see Fig. \ref{fig:model4snap}).  

In \cite{Delfau2016} the aggregation dynamics of this system was described with particle simulations and with the deterministic DKTE, which   explains them roughly. In the present section we aim, as for the previous models, to analyze what is the role of fluctuations in the density description (\textit{i.e.} in the DKTE). 

The $N$ particles evolve following 
\begin{align}
\frac{d \bx_i}{dt} =- \sum_{j=1}^{N} \boldsymbol\nabla_{\bx_i} V(\bx_i -\bx_j) + \sqrt{2D} {\boldsymbol \eta}_i(t), \label{eq:md4}
\end{align}
so that there is a force between particle $i$ and $j$: ${\bf F}_{ij} = -\boldsymbol\nabla V (\bx_i -\bx_j)$ deriving from the potential $V$, and the density dynamics is ruled by the DK equation  \cite{Dean1996,Kawasaki1994}: 
 \begin{align}
	\frac{\partial\rho(\bx,t )}{\partial t} =&  \boldsymbol\nabla \cdot 
	(\sqrt{2 D \rho(\bx,t)}~{\boldsymbol\zeta}(\bx,t) ) +D\boldsymbol\nabla^2\rho(\bx,t)
	\nonumber\\
	&+ \boldsymbol\nabla \cdot \bigg(\rho(\bx,t) \int d\by\boldsymbol\nabla V(\bx-\by)\rho(\by,t) \bigg) .
	\label{eq:Dean}
\end{align}
In the above formulae, ${\boldsymbol \eta}_i(t)$ and $\boldsymbol\zeta(\bx,t)$ have the same meaning and correlations than 
in the previous section. With these correlations, the probability of the fluctuating configurations $\rho(x,t)$ approaches an equilibrium Gibbs distribution at long times \cite{Dean1996,Kawasaki1994,Archer2004}. We use the following soft-core repulsive interacting potential between particles \cite{Delfau2016} :
\begin{align}
V (\br)  = \epsilon \exp{\bigg(-\bigg(\frac{|\br|}{R}\bigg)^\alpha\bigg)}, \label{eq:vG3}
\end{align}
where $\br = \bx_i - \bx_j$ is the relative position, $R$ is the interaction radius, $\epsilon$ is the strength of the potential and $\alpha$ gives the steepness of the kernel decay. In terms of dimensionless variables ($\tilde \bx= \bx/R$, $\tilde t = t\rho_0 \epsilon$ and $\tilde \rho=\rho/\rho_0$): 
\begin{align}
\partial_{t} {\rho}(\bx,{t}) &= \boldsymbol\nabla\cdot \bigg( {\rho}(\bx,{t}) \int d\by\;{\boldsymbol\nabla}\tilde{V}({\bx} -{\by}) {\rho}({\by},{t})\bigg) 
\nonumber	\\
	&+ \tilde{D}\boldsymbol{\nabla}^2 \rho({\bx},{t})+\frac{1}{\sqrt{n_R}} \boldsymbol{\nabla} \bigg( \sqrt{2\tilde{D}  {\rho}(\bx,{t}) } \; {\zeta}(\bx,{t})  \bigg) \ ,
	\label{eq:DK_normalized}
\end{align}
where $n_R= \rho_0 R^d$, $\tilde V(\tilde\bx) = V(\bx)/\epsilon$, $\tilde{D} = D/(\epsilon\rho_0 R^d)$, and for simplicity we have omitted all tildes except the ones over $\tilde V$ and $\tilde D$. 

Once the parameter $\alpha$ is fixed 
so that patterns may form (i.e. $\alpha>2$ \cite{Delfau2016}),
 $\tilde D$ will be taken as the main control parameter that, when decreasing, will bring the system from a homogeneous to a spatially periodic configuration. As in the previous model, we focus on the role of fluctuations on this spatial-pattern transition.

\subsection{Numerical results}
\label{Subsec:NMmodel4}

From now on we fix the parameters to $\epsilon=0.0333, R=0.1, \alpha=3$, and perform numerical simulations with the algorithm described in the Appendix. Fig.\,\ref{fig:model4snap} shows the 
long-time density configurations 
for $N = 20000$. In Fig.\,\ref{fig:patterninteract} we report the height of the maximum of the structure function under the various scenarios for two 
different numbers of particles 
(upper plot is for $N=10000$ and bottom one for $20000$). 

The impact of fluctuations is very relevant, 
going farther than shifting the critical point 
of the pattern-forming transition. When 
initializing the simulation on a homogeneous density 
with small perturbations added to it, we observe that the deterministic version of the model (black line) 
exhibits an abrupt transition at $\tilde D_c=0.082$, a value of the diffusivity below which large amplitude periodic hexagonal patterns are formed. This critical value 
can be obtained from linear analysis of the homogeneous solution in the deterministic framework \cite{Delfau2016}. A similar transition occurs for the particle simulations (red line) and
 for the DKTE, but they occur at larger values of $\tilde D$. 
Again, fluctuations anticipate the transition to patterns. If 
instead of starting at a perturbed homogeneous 
configuration, the initial condition is
a developed hexagonal pattern obtained 
at small $\tilde{D}$, and a numerical continuation 
of the solution is done by increasing $\tilde{D}$ 
and using as initial condition the final state obtained at a slightly 
lower value, the scenario changes. All the frameworks (particles, DKTE and deterministic DKTE) show a range of bistability, \textit{i.e.}, exhibits a typical hysteresis loop. The largest hysteresis loop occurs for the deterministic case. The hysteresis loops from particles and DKTE simulations approach each other for sufficiently large $N$ (and also widens to approach the deterministic case for still much larger $N$) whereas they shrink and become more dissimilar for smaller $N$.  
In general, when comparing the $N=10000$ and $N=20000$ plots we observe, as expected, that the influence of fluctuations decreases as the number of particles increases, leading to a more accurate approximation of the deterministic version. 


\begin{figure}[hbt!]
	\begin{center}
		\includegraphics[width=\columnwidth]{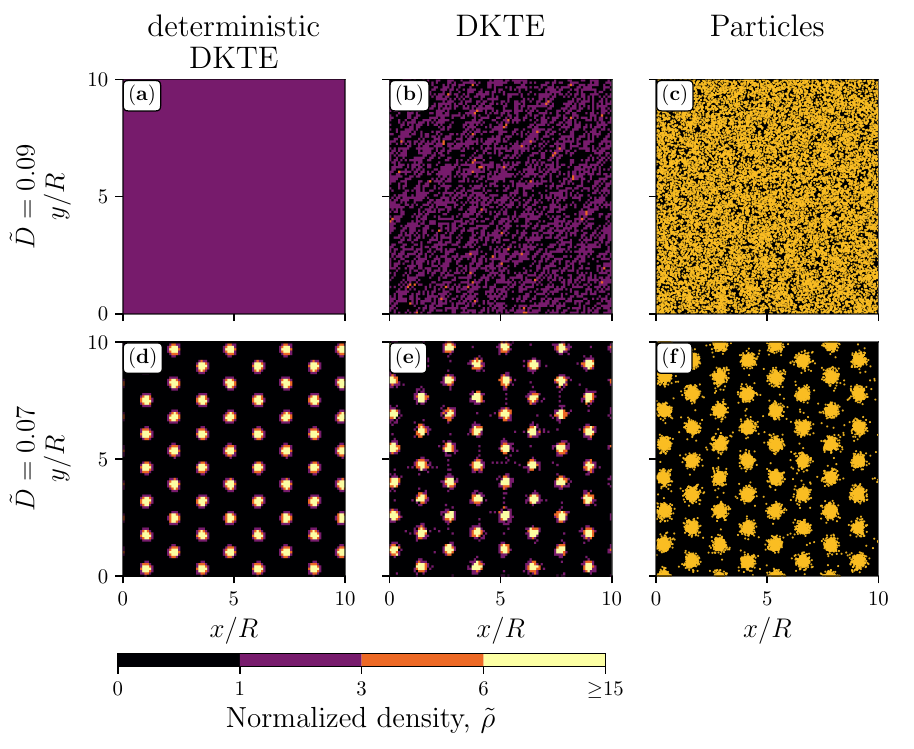}
		\caption{Steady state distributions at long times of the normalized density $\tilde\rho=\rho/\rho_0$ for Model IV and $L=1$, $R=0.1$, $\epsilon=0.0333$ and $\alpha=3$, for two different values of normalized diffusivity $\tilde D$. \textbf{(a-c)}: statistically homogeneous distributions with $\tilde D = 0.09$. \textbf{(d-f)}:  periodic density patterns for $\tilde D = 0.07$. The deterministic DKTE (\textbf{a,d}) and the DKTE (\textbf{b,e}) profiles were obtained from simulation of Eq.\,\eqref{eq:DK_normalized} with $dt = 10^{-3}, dx =10^{-2}$, and $N=\infty$ and $N=20000$, respectively. Particle data (\textbf{c,f}) where obtained from the simulations of Eq.\,\eqref{eq:md4} for $N = 20000$ and $dt = 10^{-5}$. The Particle images are shown without any spatial binning, i.e. each yellow dot indicates the position of a particle.}
		\label{fig:model4snap}
	\end{center}
\end{figure}

\begin{figure}[hbt!]
\begin{center}
\includegraphics[width=\columnwidth]{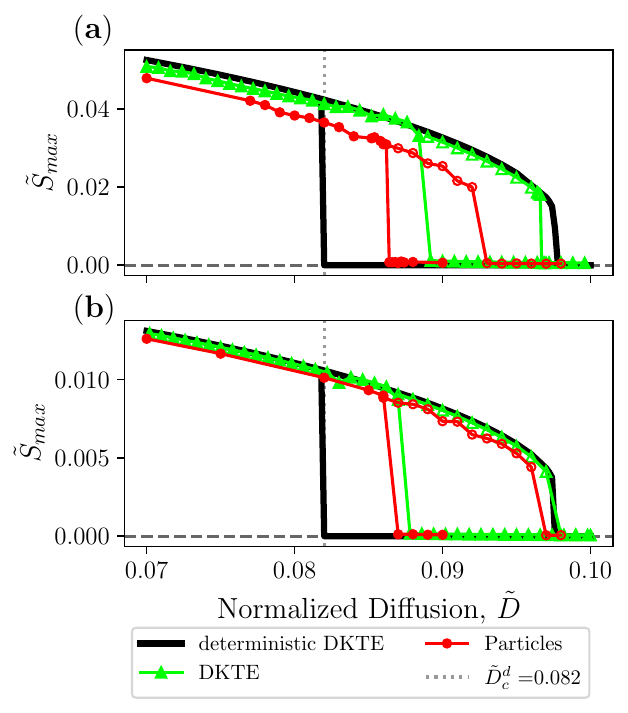}
\caption{Maximum height of the structure function normalized by the number of particles. \textbf{a}) Upper plot
is for $N=10000$ and \textbf{b}) bottom one for $N=20000$. Dashed vertical gray line indicates the transition from homogeneous to periodic
pattern as predicted by linear stability analysis of the homogeneous solution in the deterministic DKTE. Black lines give the results from simulations of the deterministic DKTE. Green line with triangles is from DKTE. Red line with circles is from particle simulations. Simulations of deterministic DKTE and DKTE were performed up to $3\times 10^{5}$ and $5\times 10^{5}$ simulation steps, respectively, while up to $10^6$ steps for the particles. Other parameters as in Fig. \ref{fig:model4snap}}. 
\label{fig:patterninteract}
\end{center}
\end{figure}


Thus, we observe that fluctuations impact in two important ways: first, there is a shift in the bifurcation point, so that spatial patterns emerge at larger values of the control parameter than predicted by mean-field/deterministic theory. Second, fluctuations shrink the bistable regime, thus reducing the strength of the hysteresis.

\section{Summary and conclusions}
\label{Sec:summary}

In this article we have investigated how fluctuations affect particle systems described by Dean-Kawasaki-type equations, which model the density dynamics of Brownian particle systems. We have explored a series of models of increasing complexity and compare their behavior for three modeling frameworks: microscopic particle simulations, DKTE, and deterministic DKTE. The models we have explored and the main impact of fluctuations were: a) Spatially-dependent diffusivity, where diffusivity varies with position. Here fluctuations introduce roughness in the density profile but, because of the linearity of the average equations, they do not alter the expected average behavior. b) Locally density-dependent diffusivity: In situations in which sharp fronts form, their speed increases due to fluctuations in contrast to other models in the literature where (non-conserved) noise typically slows fronts down.  We have argued that this effect can be understood in terms of an enhanced effective diffusivity. c) Nonlocal density-dependent diffusivity: Fluctuations advance the onset of pattern formation. And d) particles with direct repulsive soft-core interactions: The deterministic model shows hysteresis. The stochastic ones (particles and DKTE) also, 
but with a smaller width that is further reduced
 as the number of particles $N$ decreases. 
In general, particle and DKTE numerical-integration results show a similar effect of fluctuations for the same value of $N$, although some quantitative differences are seen between them, a consequence of the finite resolution of the DKTE simulations. 

These findings demonstrate that fluctuations can have nontrivial effects, emphasizing the importance of stochastic modeling in understanding collective particle dynamics. While this is a well-established message, our work specifically confirms it in the context of conservative multiplicative fluctuations, whose impact is often overlooked due to the mathematical and computational challenges they present. Future work will consider the dependence of the sharp front velocity on $N$, explore further types of interactions between particles, and investigate alternative forms of mobility such as active motion and superdiffusion.

\acknowledgments

N.S. and E.H-G. acknowledge 
funding by the Spanish Ministerio de Ciencia, 
Innovaci\'on y Universidades
(MICIU/AEI/10.13039/501100011033) through
the Maria de Maeztu project CEX2021-001164-M.
C.L. and E.H-G. acknowledge grant LAMARCA PID2021-123352OBC32
funded by MCIN/AEI/10.13039/501100011033 and FEDER, UE.


\appendix
\section{Numerical methods}
\subsection{Models I  and II}

The numerical integration of the onedimensional Eq.\eqref{DKmodel1adim}
 and Eq.\eqref{DK2adim} is performed using a single step stochastic 
Euler method  with a finite-differences scheme. The 
treatment of the multiplicative conserved noise 
follows \cite{Cornalba2026,Cornalba2023b,Wehlitz2025}, 
where mathematical results on the validity and accuracy of the method can be found. 

We approximate the continuum function $\rho(x,t)\approx \rho_j^n$, where spatial positions are discretized with a spatial resolution $h>0$, so that $x=h~j$, and the simulation time $t$ is split into equal intervals of duration $dt$, so that $t=n dt$. 
The discrete scheme is given by,
\begin{eqnarray}
\rho_j^{n+1} = \rho_j^n + dt \, f(\rho_j^n) + \mathrm{d}B(\rho_j^n),
\end{eqnarray}
with 
\begin{align}
f(\rho_j^n) = \nabla^2 \lbrack D_j \,\rho_j^n \rbrack , 
\end{align}
where $\nabla^2$ is the discrete Laplacian, and the noise increment is,
\begin{align}
\mathrm{d}B(\rho_j^n)=\frac{\sqrt{\mathrm{d}t}}{\sqrt{N h}}\nabla\cdot\bigg\lbrack \sqrt{2D_j\, \rho_{j+}^n}\,g_j^{n}\bigg\rbrack.
\end{align}
$\nabla$ is the discrete derivative. The density inside the square root in the noise term is forced to be
positive $\rho^n_{j+} \equiv \mathrm{max}\{0,\rho_j^n\}$.  $g_j^n$ is a Gaussian random number of zero mean and unit variance, independent of the other ones occurring at different positions $j$ and times $n$. 
$N$ is the number of particles and \(D_j\) is the discretization of the continuum diffusion coefficient \(D(x)\) for 
model I (Sect.\ref{Sec:model1}), and $D(\rho_j^n)$ for model II (Sect. \ref{Sec:model2}). 
The discrete derivative and Laplacian operators are computed with second order 
centered differences.

\subsection{Models III and IV}

For the non-local interaction models, models III and IV, in two dimensions
with periodic boundary conditions, we choose to perform the discretization in the Fourier version of the DKTE. This pseudospectral method of integration is useful for handling the non-local interactions improving the simulation time required for each run. It consists in performing a spatial Fourier transform of the density field and integrate it at each time step in Fourier space. Products occurring in nonlinear terms are computed in real space and then transformed back to Fourier space. The noise term is treated as before, \textit{i.e.}, making the density equal to $0$ when it gets numerically negative values. 

We use a simple Euler algorithm 
 to integrate a time step (we show explicitly 
 the computations for model IV, \textit{i.e.} for the original DK equation): 
\begin{eqnarray}
	\hat\rho_k^{n+1} = (1 - D\, dt\, k^2) \hat\rho_k^{n}  
+ \mathrm{d}\mathcal{N}\lbrack \hat\rho_k^n\rbrack + \mathrm{d}\mathcal{B}\lbrack\hat\rho_k^n\rbrack .
\end{eqnarray}
$\hat\rho_k^{n} = \mathcal{F} \lbrack\rho_j^n \rbrack$ is the Fourier transform of the discrete density at time step $n$. $\mathrm{d}\mathcal{N}\lbrack \hat\rho_k^n\rbrack$ is the non-local interaction term, and $\mathrm{d}\mathcal{B}\lbrack \hat\rho_k^n\rbrack$ is the multiplicative noise term of the DK equation. The non-local term is represented in Fourier space as
\begin{eqnarray}
\mathrm{d}	\mathcal{N}\lbrack\hat\rho_k^n\rbrack = -i\vec{k} \cdot \mathcal{F}\bigg\lbrack \rho_j^n \,\mathcal{F}^{-1}\lbrack -i \vec{k}
	\hat{V}_k\hat\rho_k^n \rbrack \bigg\rbrack \mathrm{d}t,
\end{eqnarray}
 $\hat{V}_k$ is the Fourier transform of the potential, and the noise term was integrated as
\begin{equation}
	\mathrm{d}\mathcal{B}\lbrack\hat\rho_k^n\rbrack = -i \vec k \cdot \,\sqrt{\frac{2D dt}{N dx dy}} \,\mathcal{F}\bigg\lbrack\sqrt{\rho_{j+}^n} \; \vec{g}_j^{~n}\bigg\rbrack ,
\end{equation}
where $\vec{g}_j^{~n}$ is a random vector with components sampled from a Gaussian distribution of zero mean, variance one, and independence for different sites $j$, times $n$ and vector components. We use an isotropic discretization, $dx=dy=h$. The factors in the noise-term discretization arise from the fact that for the continuous models the noise $\zeta(\bx,t)$ is a distribution that satisfies:
\begin{eqnarray}
	\left< \zeta(\bx,t)\zeta(\bx',t') \right> = \delta(\bx-\bx')\delta(t-t').
\end{eqnarray}

Concerning the particle simulation, they are also performed using
an Euler algorithm, and
the only requirement is to choose a time step $dt$ small enough such that $dt \ll t_C$, where $t_C$ is the characteristic time for a particle to travel a distance $R$ (the interaction range) \cite{Espanol1995}. 

\bibliography{bugsrefs.bib}

\end{document}